\documentclass[aps,prl,twocolumn,superscriptaddress,reprint]{revtex4-2}
\usepackage{graphicx}
\usepackage{amssymb, amsmath}
\usepackage{booktabs}
\usepackage[usenames]{color}
\usepackage{bm}
\usepackage{multirow}
\usepackage{dcolumn}
\usepackage{hyperref}
\usepackage{enumerate}
\usepackage[mathlines]{lineno}
\hypersetup{
    colorlinks=true,
    linkcolor=blue,
    filecolor=gray,      
    urlcolor=blue,
    citecolor=blue,
}

\begin{document}

\title{Efficient hybrid density functional calculation by deep learning}

\affiliation{State Key Laboratory of Low Dimensional Quantum Physics and Department of Physics, Tsinghua University, Beijing, 100084, China}
\affiliation{Tencent Quantum Laboratory, Tencent, Shenzhen, Guangdong 518057, China}
\affiliation{Institute for Advanced Study, Tsinghua University, Beijing 100084, China}
\affiliation{Beijing National Laboratory for Condensed Matter Physics, Institute of Physics, Chinese Academy of Sciences, Beijing 100190, China}
\affiliation{Songshan Lake Materials Laboratory, Dongguan, Guangdong 523808, China}
\affiliation{School of Physics, Peking University, Beijing 100871, China}
\affiliation{Key Laboratory of Quantum Information, University of Science and Technology of China, Hefei, Anhui, 230026, China}
\affiliation{Institute of Artificial Intelligence, Hefei Comprehensive National Science Center, Hefei, Anhui, 230026, China}
\affiliation{College of Chemistry and Molecular Engineering, Peking University, Beijing 100871, China}
\affiliation{Frontier Science Center for Quantum Information, Beijing, China}
\affiliation{RIKEN Center for Emergent Matter Science (CEMS), Wako, Saitama 351-0198, Japan}
\affiliation{These authors contributed equally}

\author{Zechen \surname{Tang}}
\affiliation{State Key Laboratory of Low Dimensional Quantum Physics and Department of Physics, Tsinghua University, Beijing, 100084, China}
\affiliation{These authors contributed equally}

\author{He \surname{Li}}
\affiliation{State Key Laboratory of Low Dimensional Quantum Physics and Department of Physics, Tsinghua University, Beijing, 100084, China}
\affiliation{Tencent Quantum Laboratory, Tencent, Shenzhen, Guangdong 518057, China}
\affiliation{Institute for Advanced Study, Tsinghua University, Beijing 100084, China}
\affiliation{These authors contributed equally}

\author{Peize \surname{Lin}}
\affiliation{Beijing National Laboratory for Condensed Matter Physics, Institute of Physics, Chinese Academy of Sciences, Beijing 100190, China}
\affiliation{Songshan Lake Materials Laboratory, Dongguan, Guangdong 523808, China}
\affiliation{These authors contributed equally}

\author{Xiaoxun \surname{Gong}}
\affiliation{State Key Laboratory of Low Dimensional Quantum Physics and Department of Physics, Tsinghua University, Beijing, 100084, China}
\affiliation{School of Physics, Peking University, Beijing 100871, China}

\author{Gan \surname{Jin}}
\affiliation{Key Laboratory of Quantum Information, University of Science and Technology of China, Hefei, Anhui, 230026, China}

\author{Lixin \surname{He}}
\affiliation{Key Laboratory of Quantum Information, University of Science and Technology of China, Hefei, Anhui, 230026, China}
\affiliation{Institute of Artificial Intelligence, Hefei Comprehensive National Science Center, Hefei, Anhui, 230026, China}

\author{Hong \surname{Jiang}}
\affiliation{College of Chemistry and Molecular Engineering, Peking University, Beijing 100871, China}

\author{Xinguo \surname{Ren}}
\email{renxg@iphy.ac.cn}
\affiliation{Beijing National Laboratory for Condensed Matter Physics, Institute of Physics, Chinese Academy of Sciences, Beijing 100190, China}
\affiliation{Songshan Lake Materials Laboratory, Dongguan, Guangdong 523808, China}

\author{Wenhui \surname{Duan}}
\email{duanw@tsinghua.edu.cn}
\affiliation{State Key Laboratory of Low Dimensional Quantum Physics and Department of Physics, Tsinghua University, Beijing, 100084, China}
\affiliation{Tencent Quantum Laboratory, Tencent, Shenzhen, Guangdong 518057, China}
\affiliation{Institute for Advanced Study, Tsinghua University, Beijing 100084, China}
\affiliation{Frontier Science Center for Quantum Information, Beijing, China}

\author{Yong \surname{Xu}}
\email{yongxu@mail.tsinghua.edu.cn}
\affiliation{State Key Laboratory of Low Dimensional Quantum Physics and Department of Physics, Tsinghua University, Beijing, 100084, China}
\affiliation{Tencent Quantum Laboratory, Tencent, Shenzhen, Guangdong 518057, China}
\affiliation{Frontier Science Center for Quantum Information, Beijing, China}
\affiliation{RIKEN Center for Emergent Matter Science (CEMS), Wako, Saitama 351-0198, Japan}

\begin{abstract}

Hybrid density functional calculation is indispensable to accurate description of electronic structure, whereas the formidable computational cost restricts its broad application. Here we develop a deep equivariant neural network method (named DeepH-hybrid) to learn the hybrid-functional Hamiltonian from self-consistent field calculations of small structures, and apply the trained neural networks for efficient electronic-structure calculation by passing the self-consistent iterations. The method is systematically checked to show high efficiency and accuracy, making the study of large-scale materials with hybrid-functional accuracy feasible. As an important application, the DeepH-hybrid method is applied to study large-supercell Moir\'{e} twisted materials, offering the first case study on how the inclusion of exact exchange affects flat bands in the magic-angle twisted bilayer graphene.
\end{abstract}

\maketitle

A milestone development of density functional theory (DFT) is the invention of hybrid functionals, developed first as an ad hoc correction to local density or generalized gradient  approximations (LDA/GGA)~\cite{hybrid}, and later formulated more rigorously in the generalized Kohn-Sham framework~\cite{gks}. Superior to conventional density functionals, hybrid functionals provide a viable route to solve the critical ``band-gap problem'' of DFT~\cite{perdew1985,Perdew2017}, thus indispensable for reliable material prediction and particularly useful for comoutational studies in (opto-)electronics, spintronics, topological electronics, etc. The practical use of hybrid functionals, however, is seriously limited, because their computational cost is considerably higher than local and semilocal DFT methods. Great efforts have been devoted to improving the numerical algorithms~\cite{
hybrid_algo1982,hybrid_algo1989,hybrid_algo1996,hybrid_algo2009,hybrid_algo2010,hybrid_algo2012,hybrid_algo2015,hybrid_algo2020,abacus_hybrid}. This helps reduce the computational overhead and facilitates linear-scaling hybrid-functional calculations, but cannot fundamentally change the landscape of \textit{ab initio} computations.

Deep learning methods shed light on revolutionizing \textit{ab initio} materials simulation
~\cite{LORENZ2004,Carleo2019,Behler2007,Zhang2018,Gilmer2017,Schutt2018,jorgensen2018,Xie2018,Schutt2019,Anderson2019,Unke2021_2,Li2022,Li2022_2,Gong2022,xDeepH2022,Klicpera2020,Unke2021,Gu2022,Su2022,Zhong2022,Batzner2022,Musaelian2022,Qiao2022,nigam2022,zhang2022}. For instance, the use of artificial neural networks to represent DFT Hamiltonian enables efficient electronic-structure calculations with \textit{ab initio} accuracy, whose computational cost is as low as that of empirical tight-binding calculationss~\cite{Li2022,Li2022_2,Gong2022,xDeepH2022}. The so-called deep-learning DFT Hamiltonian (DeepH) approach has been demonstrated powerful in large-scale materials simulation for both nonmagnetic and magnetic systems~\cite{Li2022,Li2022_2,Gong2022,xDeepH2022}. However, the method was originally designed within the Kohn-Sham (KS) framework. Therein the deep-learning problem is simplified by the local nature of the exchange-correlation potential. In contrast, hybrid functionals are usually done within the generalized Kohn-Sham (gKS) scheme~\cite{gks}, giving rise to non-local exchange potentials. Considering that the deep-learning approach relies critically on the locality property~\cite{Unke2021_3,Chen2021,Zepeda2021}, whether the same strategy is applicable to the generalized Kohn-Sham scheme or not is an important open question.

In this work, we find that the gKS-DFT Hamiltonian of hybrid functionals $H_{\text{DFT}}^{\text{hyb}}$ can be represented by neural networks as well as for conventional DFT, benefiting from the preservation of nearsightedness principle in the localized basis. We apply deep E(3)-equivariant neural networks to model $H_{\text{DFT}}^{\text{hyb}}$ as a function of material structure. The method is tested to show good performance by systematic numerical experiments and further applied to study Moir\'{e}-twisted superstructures, such as magic-angle twisted bilayer graphene, demonstrating the capability for large-scale electronic-structure calculations with hybrid-functional accuracy. Our work paves the way for accurate, efficient materials simulation, and also opens a door for developing deep-learning \textit{ab initio} methods beyond DFT.

In the KS-DFT~\cite{Kohn1965}, the challenging interacting-electron problem is mapped to an auxiliary non-interacting  problem whereby the complicated many-body effects are incorporated in an exchange-correlation functional. Within the conventional approximations of KS-DFT, the exchange-correlation energy is expressed as an {\it explicit functional of  density} and a {\it local} form of exchange-correlation potential $V_{\text{xc}}(\mathbf{r})$ is assumed. Such approximations greatly simplify the problem and is widely used in \textit{ab initio} calculations. Unfortunately, the self-interaction error is prevalent in such density-based functionals, which could result in systematic failures of DFT, including the band gap problem~\cite{perdew1985,cohen2008}.
In fact, even with the exact exchange-correlation functional, the fundamental band gap will still be underestimated within the KS-DFT framework
\cite{Perdew1982,Perdew1983}. This critical issue needs to be addressed to make reliable property prediction on electronic materials.

\begin{figure*}[t]
    \centering
    \includegraphics[width=0.8\linewidth]{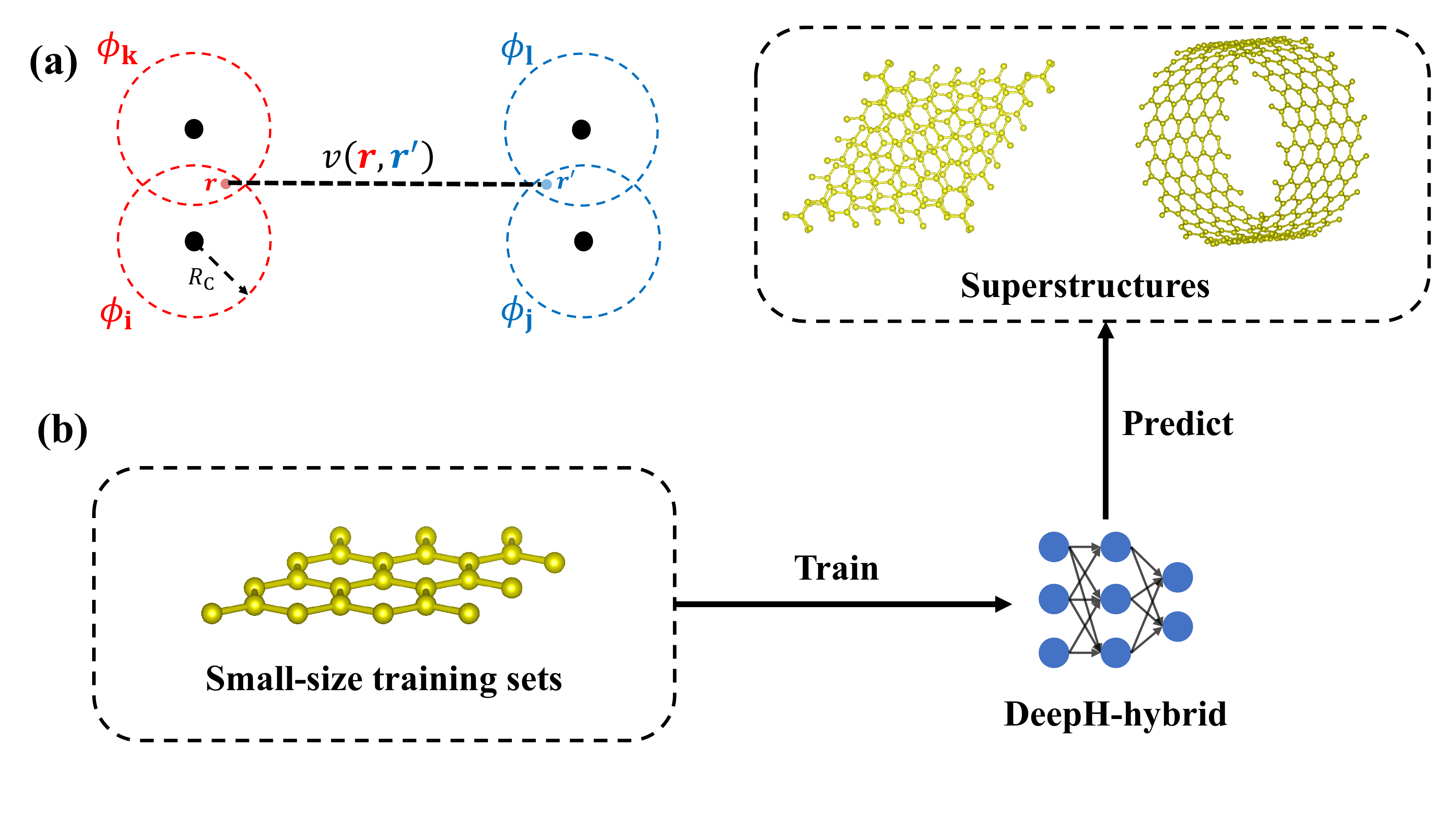}
    \caption{(\textbf{a}) Non-local nature of exact exchange potential. Two non-overlapping atomic orbitals $\phi_{\textbf{i}}$ and $\phi_{\textbf{j}}$ has a non-zero Hamiltonian matrix element due to the presence of non-local potential $v(\mathbf{r},\mathbf{r}')$. Other atomic orbitals $\phi_{\textbf{k}}$ and $\phi_{\textbf{l}}$ are involved in the non-local coupling. (\textbf{b}) Schematic workflow of the deep-learning method named DeepH-hybrid.} 
    \label{fig:1-workflow-nearsightedness.pdf}
\end{figure*}

The generalized Kohn-Sham scheme allows the use of orbital-dependent exchange-correlation potential, which helps relieve the band-gap problem. As a typical example, hybrid-functional methods replace portion of semi-local exchange with the (screened) Hartree–Fock exact exchange, by which the band gap problem could be largely resolved. However, a non-local, exact-exchange potential $V_{\text{Ex}}(\mathbf{r},\mathbf{r}')$ will be introduced into the effective Hamiltonian, which significantly complicates the calculation. Let us illustrate this with the localized orbital basis functions $\phi_\mathbf{i}(\mathbf{r}) = R_{ipl}(r)Y_{lm}(\hat{r})$, where $R_{ipl}$ is the radial function centered at the $i_{\text{th}}$ atom labelled by multiplicity $p$ and angular momentum quantum number $l$, $Y_{lm}$ is the real spherical harmonics of degree $l$ and order $m$, and $\mathbf{i} \equiv (iplm)$ is used for simplicity. In the localized basis, the Kohn-Sham eigenstate is $\psi_n(\mathbf{r}) = \sum_\mathbf{i} c_{n\mathbf{i}} \phi_\mathbf{i}(\mathbf{r})$, and the (screened) exact-exchange potential is written as
\begin{align}
\label{V_EX}
V^{\text{Ex}}_{\mathbf{ij}} &=-\sum_{n}^{\text{occ}} \sum_{\mathbf{k,l}}  c_{n\mathbf{k}} c^*_{n\mathbf{l}} (\mathbf{ik}|\mathbf{lj}), \\ \nonumber 
(\mathbf{ik}|\mathbf{lj}) &=  \int \int d\mathbf{r} d\mathbf{r'}\phi_\mathbf{i}^*(\mathbf{r})\phi_\mathbf{k}(\mathbf{r}) v(\mathbf{r}-\mathbf{r'})
\phi^*_\mathbf{l}(\mathbf{r'})\phi_\mathbf{j}(\mathbf{r'}),
\end{align}
where $v(\mathbf{r}-\mathbf{r'})$ denotes the Coulomb potential $1/|\mathbf{r}-\mathbf{r'}|$ or its screened version. Note that the two-electron Coulomb repulsion integral $(\mathbf{ik}|\mathbf{lj})$ involves a four-center integration over six spatial coordinates (Fig.~\ref{fig:1-workflow-nearsightedness.pdf}a), and the number of integrals to be calculated is enormous, growing quickly with the system size. Hence the computation becomes much more expensive than local or semilocal DFT. The situation is alleviated  with significant algorithm improvements (e.g., resolution of identity and linear scaling techniques) \cite{levchenko2015hybrid,hybrid_algo2020,abacus_hybrid}, but the significant increase of the computational cost from semilocal to hybrid 
DFT methods is not fundamentally changed. This is the major drawback of hybrid functionals, which restricts broad applications of the methods.

The exchange-correlation potentials of hybrid functionals share the form $V_{\text{xc}}^\text{hyb}(\mathbf{r},\mathbf{r}') = V_{\text{xc}}'(\mathbf{r})\delta(\mathbf{r}-\mathbf{r}') + \alpha V_{\text{Ex}}(\mathbf{r},\mathbf{r}')$, where a fraction $\alpha$ of the semilocal exchange potential is replaced by $V_{\text{Ex}}$ and the remaining part $V_{\text{xc}}'$ is the same as that of the semilocal DFT. For example, the Heyd–Scuseria–Ernzerhof (HSE) hybrid functional uses an error-function-screened Coulomb potential in $V_{\text{Ex}}$ with $\alpha = 25\%$~\cite{hse2003}. According to the Hohenberg-Kohn theorem~\cite{Hohenberg1964}, the auxiliary non-interacting Hamiltonian is uniquely determined by the external potential $V_{\text{ext}}$ that is defined by the material structure $\{\mathcal{R}\}$. Thus $H_{\text{DFT}}^{\text{hyb}}$ and $V_{\text{Ex}}$ can be expressed as a function of $\{\mathcal{R}\}$. It has been established that the Kohn-Sham DFT Hamiltonian $H_{\text{DFT}}^{\text{KS}}(\{\mathcal{R}\})$ can be well represented by deep neural networks~\cite{Li2022,Li2022_2,Gong2022}. We will attempt to use neural networks to model $H_{\text{DFT}}^{\text{hyb}}(\{\mathcal{R}\})$ to generalize the deep-learning approach to achieve hybrid-functional accuracy (Fig.~\ref{fig:1-workflow-nearsightedness.pdf}b). Compared with the Kohn-Sham case, a special non-local component $V_{\text{Ex}}(\{\mathcal{R}\})$ is introduced here, whose neural-network representation has not been considered before. 

Satisfying the nearsightedness principle is essential to simplify the deep-learning Hamiltonian problem, as learned from the study of KS-DFT. In the localized basis, the KS-DFT Hamiltonian can be viewed as an \textit{ab initio} tight-binding Hamiltonian. The hopping between atoms $i$ and $j$, namely the Hamiltonian matrix block $H_{ij}$, is nonzero only when the atomic distance $r_{ij}$ is smaller than a cutoff radius $R_{\text{C}}$. Moreover, $H_{ij}$ is predominately determined by the neighboring environment, whose value is insensitive to distant variations of atomic structure. Thus $H_{ij}$ can be simplified to be a function of $\{\mathcal{R}\}_{\text{N}}$, which includes structural information of neighboring atoms $\{k\}$ with $r_{ik}, r_{jk} < R_{\text{N}}$, where $R_{\text{N}}$ denotes a nearsightedness length.

For hybrid functionals, whether the non-local exact exchange is compatible with the nearsightedness principle or not should be checked. In Eq.~\ref{V_EX}, the KS eigenvector $c_{n\mathbf{k}}$ or $c_{n\mathbf{l}}$ can be influenced by distant change of boundary conditions, which breaks the nearsightedness principle. The two-electron Coulomb repulsion integral $(\mathbf{ik}|\mathbf{lj})$ displays long-distance (short-distance) decay between atoms $i$ and $j$ when the bare (screened) Coulomb potential is considered. Thus the product $c_{n\mathbf{k}} c^*_{n\mathbf{l}} (\mathbf{ik}|\mathbf{lj})$ is a non-local quantity, whose dependence on material structure is expected to be very complicated. However, the summation over occupied states $\sum_{n}^{\text{occ}} c_{n\mathbf{k}} c^*_{n\mathbf{l}}$ yields the density matrix element $\rho_{\mathbf{k,l}}$ that is a local quantity~\cite{Kohn1996,Prodan2005}. Moreover, the Coulomb integral $(\mathbf{ik}|\mathbf{lj})$ is nonzero only when both atom pairs $i$-$k$ and $l$-$j$ have finite orbital overlaps. Hence the local property gets preserved for $V^{\text{Ex}}_{\mathbf{ij}}$. This is the reminiscent of W. Kohn's principle, which states that the nearsightedness of electronic matter is originated from wave-mechanical destructive interference in many-particle systems~\cite{Kohn1996,Prodan2005}. Benefiting from the nearsightedness property,  $V^{\text{Ex}}_{\mathbf{ij}}$ can be determined by local structural information of neighborhood, similar as for local exchange-correlation potentials. The merit enables us to treat the conventional and generalized KS DFT within a unified deep-learning framework. The inclusion of exact exchange, however, could significantly weaken the sparseness and nearsightedness properties of DFT Hamiltonian. This will be accounted by changing the important length scales $R_\mathrm{C}$ and $R_\mathrm{N}$ in the design of deep neural networks.

\section{Methods}

\begin{figure*}[t]
    \centering
    \includegraphics[width=0.9\linewidth]{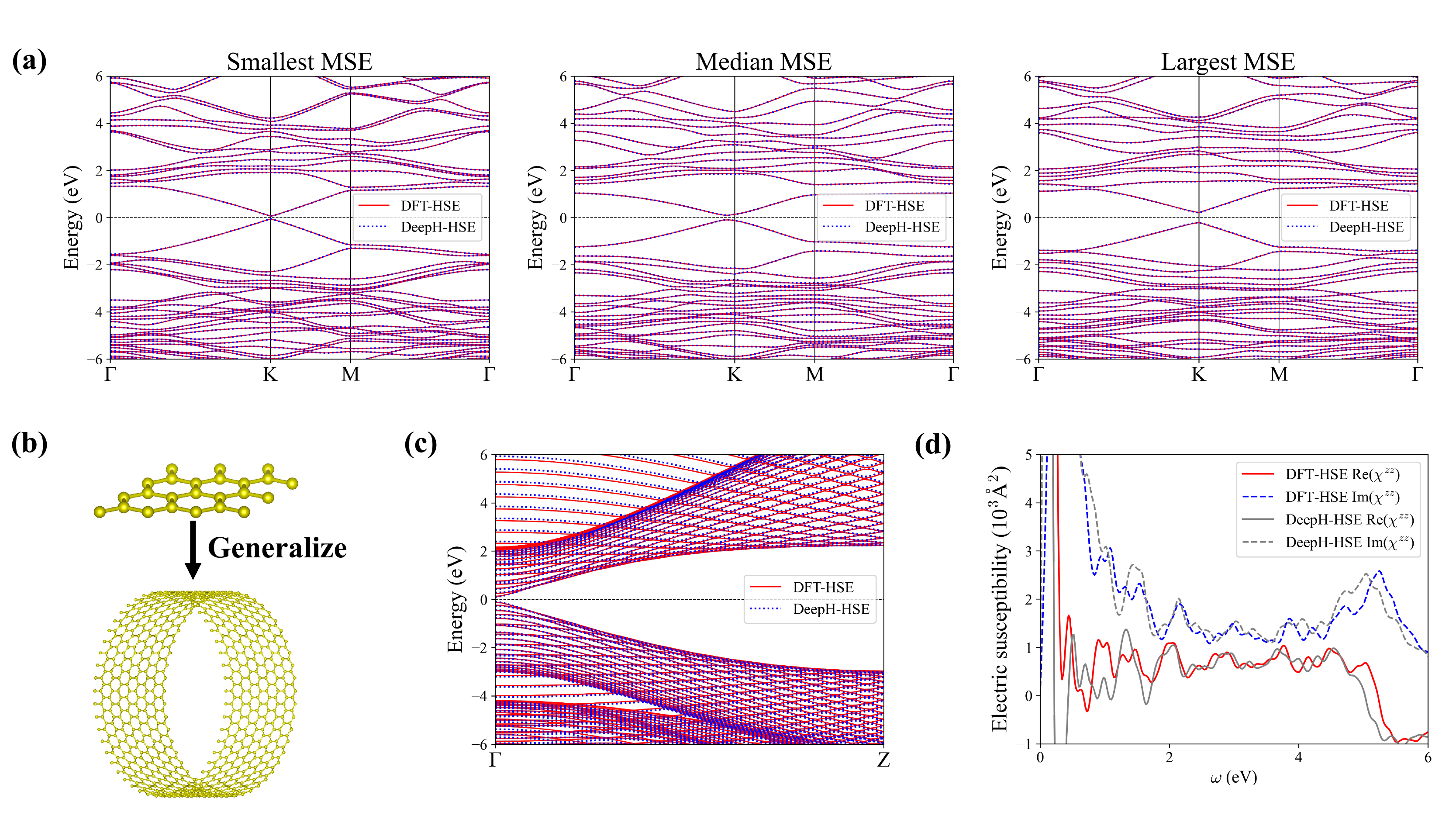}
    \caption{Example studies on monolayer graphene. (\textbf{a}) Comparison of band structures computed by DFT-HSE and DeepH-HSE for supercells of monolayer graphene. Results for test structures with the smallest, median and largest MSE are displayed. (\textbf{b}) Schematic workflow of DeepH-hybrid, which learns from training sets of nearly flat structures and is then applied to make predictions on nanotube structures. (\textbf{c,d}) Comparison of (\textbf{c}) band structures and (\textbf{d}) electric susceptibility $\chi^{zz}$ as a function of frequency $\omega$ computed by DFT-HSE and DeepH-HSE for (49, 0) carbon nanotube (CNT). The periodic direction of CNT is defined as the $z$-axis.}
    \label{fig:2-graphene.pdf}
\end{figure*}

In our work, we use the  E(3)-equivariant deep learning DFT Hamiltonian (DeepH-E3)~\cite{Gong2022} framework to model the mapping from the crystal structure $\{\mathcal R\}$ to the corresponding hybrid functional DFT Hamiltonian $H^\text{hyb}_\text{DFT}$ under numerical atomic orbital (NAO) basis. A graph is associated to each material structure with each vertex representing an atom, and edges are connected between atoms within certain cutoff. The feature vectors associated with vertices and edges are iteratively updated with neural networks, and the final edge features are the output hopping matrices. Updating a vertex or edge only uses information within its neighborhood, and so the nearsightedness property is utilized for the prediction of Hamiltonians. Moreover, since the Hamiltonian transforms covariantly between coordinate frames, it is most natural and advantageous to construct a neural network that explicitly handles the covariant property of the Hamiltonian. To achieve this, all the input, output and internal vectors of the neural network transfer according to some irreducible representations of the O(3) group under coordinate rotations and inversion. The incorporation of the requirements of locality and symmetry as \textit{a priori} knowledge has greatly enhanced the performance of DeepH-E3 and has led to its sub-meV level accuracy and excellent generalizability to large-scale materials over $10^4$ atoms.

Utilizing the equivariant neural network (ENN)~\cite{e3nn_software2022,e3nn2022}, the mapping $\{\mathcal{R}\}\to\hat{H}_{\text{DFT}}$ in DeepH-E3 is equivariant with respect to the Euclidean group in three-dimensional space ($\text{E}(3)$). E$(3)$ group is composed of translations, rotations and spatial inversion in three-dimensional space, thus containing most important symmetries in physical systems. To realize equivariance in neural network, DeepH-E3 labels each network feature with angular quantum number $l$. Upon spatial rotation $\textbf{R}$ to the input structure, all features will transform as ${\mathbf x}^{l}_{m}\stackrel{\mathbf R}{\longrightarrow}\sum_{m'} D^{l}_{mm'}(\mathbf R)\mathbf x^{l}_{m'}$, in which $D_{mm'}^{l}(\mathbf R)$ is the Wigner-D matrix. DFT Hamiltonians block $H_{ij}$ can be divided into sub-blocks $\mathbf h \equiv [H_{ij}]^{p_1p_2}$ by grouping orbitals with the same $p$ together. The resulting sub-blocks are equivariant tensor whose elements transform as $ \textbf{h}_{m_1m_2}^{l_1l_2}\to \sum_{m_1',m_2'}D^{l_1}_{m_1m_1'}(\textbf{R})D^{l_2}_{m_2m_2'}\textbf{h}_{m_1'm_2'}^{l_1l_2}$ upon rotation $\textbf{R}$. Equivariant tensors and equivariant vectors can be associated by Wigner-Eckart theorem $l_1 \otimes l_2 = |l_1-l_2|\oplus \cdots \oplus (l_1+l_2)$. Hamiltonian sub-blocks are regarded as equivariant tensors with representation $l_1\otimes l_2$ and constructed accordingly. DeepH-E3 starts with $l=0$ (scalar) features by embedding atomic numbers ($Z_i$) and distance between atom pairs ($|\mathbf{r}_{ij}|$). Relative direction of atom pairs are also taken as input features with $l=1,2,\cdots$ by enforcing spherical harmonics on $\hat{\mathbf{r}}_{ij}$. Regarding equivariance with respect to spatial inversion, feature vectors are additionally labelled by their parity upon spatial inversion, either even (e) or odd (o). All intermediate ENN operations are designed to preserve features' parity characteristic. 

We use the ABACUS package~\cite{abacus2016,abacus2010,abacus_hybrid} to carry out hybrid DFT calculations with norm-conserving pseudopotentials~\cite{ncpp1993} using the NAO basis. HSE06 functional with Hartree-Fock mixing constant $a=0.25$ and screening parameter $\omega=0.11$~Bohr$^{-1}$ is applied in all calculations~\cite{hse2003,hse2006}. For monolayer graphene, carbon nanotube (CNT) and bilayer graphene, C$6.0-2s2p1d$ NAOs are applied, with 13 basis functions for each carbon and cut-off radius $6.0$~Bohr. For monolayer graphene and CNT, dataset is composed of 500 random structures with $5\times5$ graphene supercell. Random structures are generated by introducing random offset up to 0.1\,\AA\,on each atom to the equilibrium configuration. For bilayer graphene, dataset is composed of 1000 random structures with $4\times4$ supercell of bilayer graphene. In addition to random offset on each atom, an overall in-plane shift is randomly assigned to each structure of bilayer graphene, and the interlayer distance is randomly sampled with normal distribution with mean 3.408\,\AA\, and standard deviation 0.047\,\AA\,. A $9\times9\times1$ Monkhorst-Pack $k$-mesh~\cite{monkhorst1976} is applied for both monolayer and bilayer graphene.

In terms of nearsightedness of Hamiltonian matrix, we set the cutoff radius $R_\mathrm{C}=24$~Bohr, which is twice the cutoff radius of Hamiltonian matrix of local functionals. This cutoff radius strikes balance between accuracy and efficiency of the Hamiltonian and well reflects the non-local nature of hybrid DFT. Datasets are randomly split into training, validation and test sets with a ratio of 6:2:2. In all trainings, our neural network is composed of three message-passing blocks, with $64\times0e+32\times1o+16\times2e+8\times3o+8\times4e$ equivariant layer for each intermediate layers. Here, $64\times0e$ stands for $64$ even-parity equivariant vectors with $l=0$, $32\times1o$ stands for $32$ odd-parity equivariant vectors with $l=1$, etc. Atomic configuration information is embedded into $64$-dimensional equivariant vectors as initial vertex and edge features. Mean squared error (MSE) of all Hamiltonian matrix elements is applied as the loss function.

\begin{figure*}[t]
    \centering
    \includegraphics[width=0.9\linewidth]{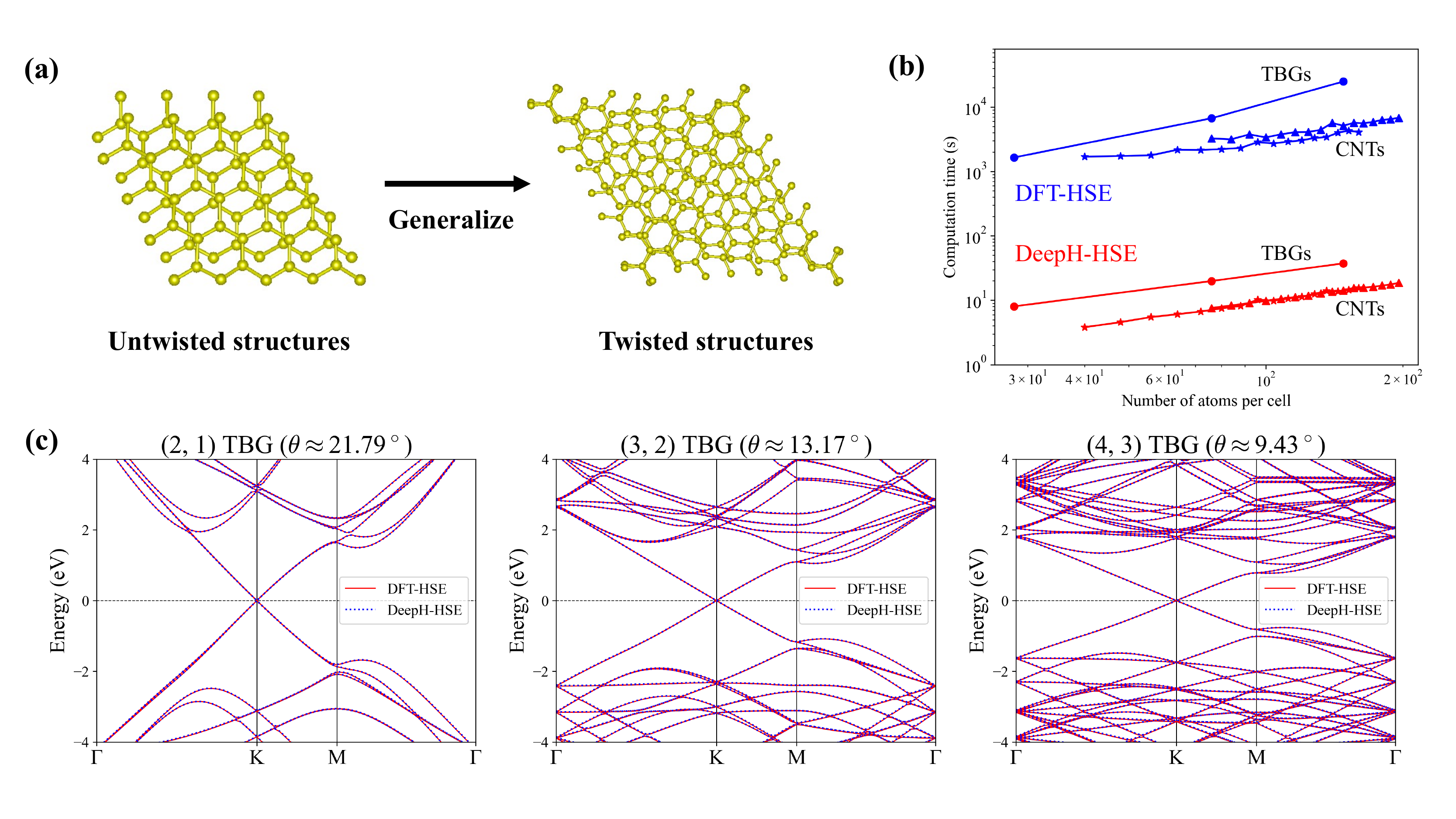}
    \caption{Example studies on twisted bilayer graphene (TBG). (\textbf{a}) Schematic workflow of DeepH-hybrid, which learns from training sets of untwisted bilayers and is then applied to make predictions on bilayers of arbitrary twist angle $\theta$. (\textbf{b}) Comparison of computation time between DFT-HSE and DeepH-HSE for example studies of TBGs and CNTs. (\textbf{c}) Band structures of (2,1), (3,2) and (4,3) TBGs computed by DFT-HSE and DeepH-HSE.}
    \label{fig:3-BG.pdf}
\end{figure*}

\section{Results}

We carry out example studies on monolayer and bilayer graphene. Figure \ref{fig:2-graphene.pdf} shows DeepH's capability on HSE Hamiltonians of monolayer graphene and generalizability to carbon nanotubes (CNT). The final mean squared error (MSE) of all Hamiltonian matrix elements is $0.415$~meV$^2$, $0.423$~meV$^2$, $0.424$~meV$^2$ for training, validation and test sets, respectively. The test set consisting of 100 perturbed graphene supercells are sorted in terms of MSE, and the band structure of the best, median and worst performing configurations are shown in Fig.~\ref{fig:2-graphene.pdf}a. In addition, we examine the model's robustness by carrying out test study on CNTs. DeepH model trained on hybrid Hamiltonians of monolayer graphene is able to predict the band structure of CNT with high accuracy, as demonstrated in Fig.~\ref{fig:2-graphene.pdf}c. One major improvement of hybrid functional is the improved accuracy of band gap, which is closely related to optical properties. We perform electric susceptibility calculation of (49, 0) zigzag CNT with HSE Hamiltonians from DFT calculation and DeepH prediction using the method developed in Ref. \cite{hoptb2019}. Real and imaginary parts of $\chi^{zz}$ as function of light frequency $\omega$ are depicted in Fig.~\ref{fig:2-graphene.pdf}. Despite minor mismatch, DeepH is sufficient to reproduce optical properties from hybrid DFT calculation.

Twisted bilayer graphene (TBG), along with other Moir\'e twisted materials are gaining increasing interest in recent years~\cite{cao2018a,cao2018b,Yankowitz2019,cao2020,Xie2021}. Magic angle TBG (MATBG) with twisting angle $\theta\approx1.08^{\circ}$ with a 11164 atom unit cell is well known for its flat bands near the Fermi energy and corresponding novel physical phenomena~\cite{flat-band2011,tarnopolsky2019,matbg-relax2019}. \textit{Ab initio} research on MATBG is heavily limited by computation cost, and the difficulty is further compounded by the larger computation cost of hybrid functional. In our previous work, DeepH is proved to be applicable to studying PBE band structure of MATBG, and thus DeepH may as well pave the way to electronic properties of MATBG with hybrid-level functionals. We examine DeepH's accuracy on twisted bilayer graphene with smaller unit cell, as displayed in Fig.~\ref{fig:3-BG.pdf}. Figure~\ref{fig:3-BG.pdf}a depicts the schematic workflow. DeepH model is first trained on perturbed supercell of untwisted bilayer graphene and then generalized to twisted structures. Each configuration in the training set are assigned a random in-plane overall shift to simulate local environment of different stackings. Interlayer distance of the dataset are also randomly sampled to enable the model to study the effect of relaxation of atomic configuration in TBG. Figure~\ref{fig:3-BG.pdf}c displays comparison of the band structure of (2, 1) (with twist angle $\theta\approx21.79^\circ$ and 28 atoms in unit cell), (3, 2) (with $\theta\approx13.17^\circ$ and 76 atoms in unit cell) and (4, 3) (with $\theta\approx9.43^\circ$ and 148 atoms in unit cell) TBG. The precise matching of band structures indicates DeepH's generalizability from untwisted bilayer graphene to TBG. The CPU time of calculating Hamiltonians with DFT and predicting with DeepH is compared in Fig.~\ref{fig:3-BG.pdf}b. DeepH method reduces the computational cost by orders of magnitude and shows a lower scaling with respect to atom numbers. Thus, DeepH is promising to study the HSE electronic property of MATBG with both efficiency and accuracy.

\begin{figure}[t]
    \centering
    \includegraphics[width=0.8\linewidth]{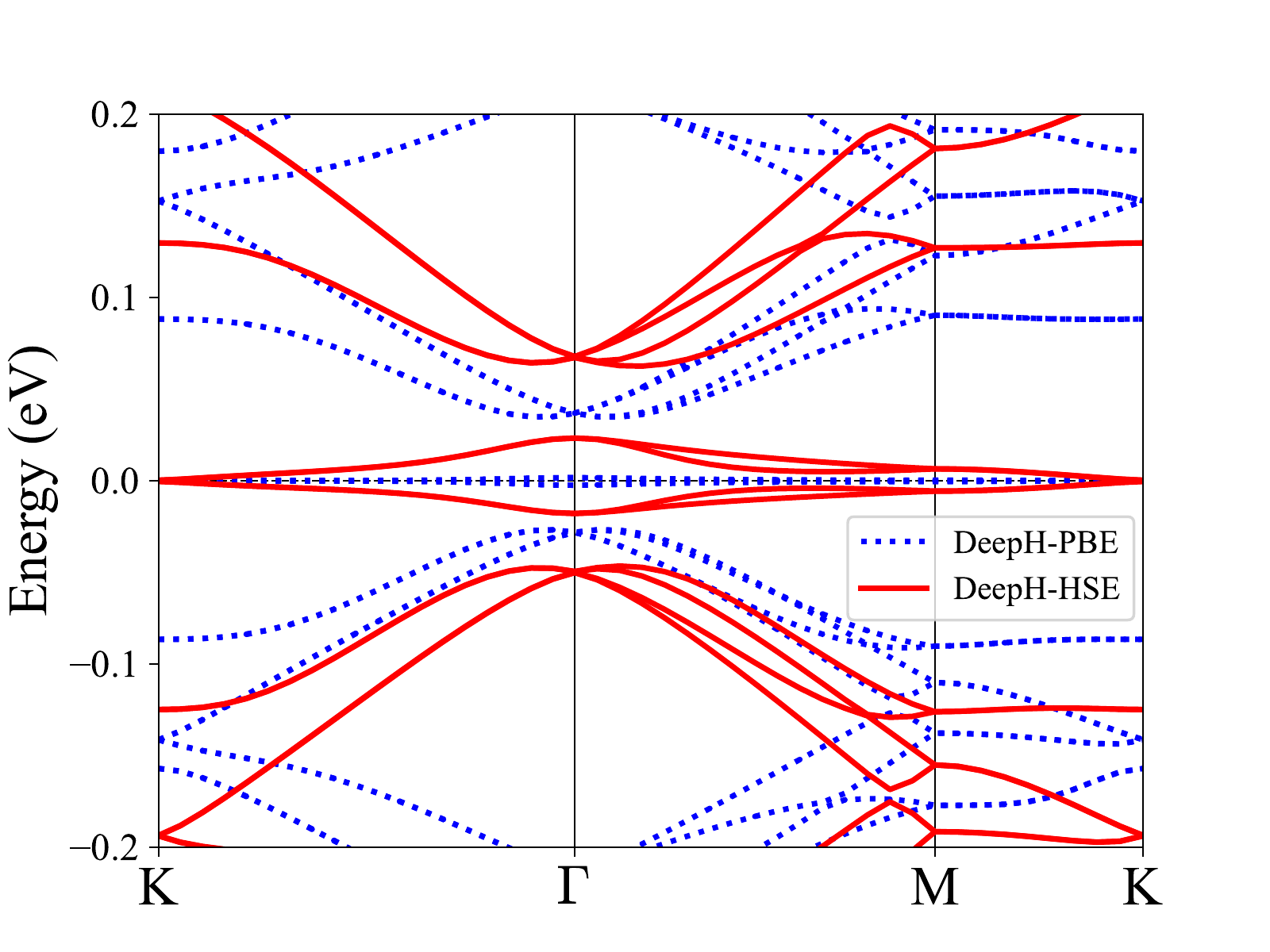}
    \caption{Band structures of magic-angle TBG ($\theta\approx1.08^\circ$, 11164 atoms per primitive cell, atomic structure relaxed by Ref.~\cite{matbg-relax2019}) computed by DeepH-PBE and DeepH-HSE. The results of DeepH-PBE are adapted from Ref.~\cite{Gong2022}, which agree well with the results computed by DFT-PBE.}
    \label{fig:4-MATBG.pdf}
\end{figure}

Figure \ref{fig:4-MATBG.pdf} demonstrates our results of PBE and HSE band structures of MATBG with structure relaxed by Ref.~\cite{matbg-relax2019}. Both PBE and HSE Hamiltonians bear four flat bands near Fermi surface. Compared with the PBE band structure, bandwidth of the flat bands of HSE band structure is increased to 41.1~meV from 4.1~meV. In addition, perturbation theory is applied to calculate the Fermi velocity of HSE and PBE Hamiltonians, yielding $v_F$=36.2~$m/s$ and 0.9~$m/s$, respectively. From our calculation, introduction of exact exchange dramatically weakens the flatness of MATBG's flat bands, and thus could have a qualitative impact on the flat band physics of MATBG.

\section{Discussion}
Owing to the preservation of nearsightedness principle, DeepH's capability is proved in predicting hybrid DFT Hamiltonians in multiple case studies. Regarding the improved accuracy of band gap with hybrid functionals, our work enables efficient study on optical properties, non-adiabatic molecular dynamics, etc., in which unoccupied conduction bands play important role. By bypassing the time-consuming self-consistent field iterations and four-center Coulomb repulsion integrals, DeepH has potential to study electronic properties of superstructures at hybrid functional level, which is previously bottlenecked due to the timely cost. Test study on MATBG reveals DeepH's ability to apply hybrid-level functionals to Moir\'{e} twisted superstructures with over $10^4$ atoms. Application of DeepH to MATBG shows a dramatic change of flat band properties when HSE functional is applied, implying that the exact exchange could have a qualitative impact on the flat band physics of MATBG. DeepH's success on HSE functionals can be a starting point for generalization of DeepH to Hamiltonians from higher-level electronic structure theory, and this methodology may overcome the accuracy-efficiency dilemma of DFT.

\section{Acknowledgments}
This work was supported by the Basic Science Center Project of NSFC (grant no. 51788104), the National Science Fund for Distinguished Young Scholars (grant no. 12025405), the Ministry of Science and Technology of China (grant nos. 2018YFA0307100 and 2018YFA0305603), the National Natural Science Foundation of China (grant nos. 12134012 and 12188101), the Beijing Advanced Innovation Center for Future Chip (ICFC), and the Beijing Advanced Innovation Center for Materials Genome Engineering.

\bibliography{reference}

\end{document}